\documentclass[aps,prc,reprint,superscriptaddress,nofootinbib,amsmath,amssymb]{revtex4-1}
\usepackage{graphicx}
\usepackage{dcolumn}
\usepackage{bm}
\usepackage{xcolor}
\usepackage{natbib}
\usepackage{url}
\usepackage{textcase}
\usepackage{bm}
\usepackage[normalem]{ulem} 
\usepackage[export]{adjustbox}

\usepackage[caption=false,font=normalsize]{subfig}
\usepackage{fixltx2e}

\begin{document}
\title{Simultaneous measurement of organic scintillator response to carbon and proton recoils}

\author{T.~A.~Laplace}
    \email[Corresponding author: ]{lapthi@berkeley.edu}
    \affiliation{Department of Nuclear Engineering, University of California, Berkeley, California 94720 USA}
\author{B.~L.~Goldblum}
    \affiliation{Department of Nuclear Engineering, University of California, Berkeley, California 94720 USA}
    \affiliation{Nuclear Science Division, Lawrence Berkeley National Laboratory, Berkeley, California 94720 USA}    
\author{J.~J.~Manfredi}
    \affiliation{Department of Nuclear Engineering, University of California, Berkeley, California 94720 USA}
\author{J.~A.~Brown}
    \affiliation{Department of Nuclear Engineering, University of California, Berkeley, California 94720 USA}
\author{D.~L.~Bleuel}
    \affiliation{Lawrence Livermore National Laboratory, Livermore, California 94550 USA}
\author{C.~A.~Brand}
    \affiliation{Department of Nuclear Engineering, University of California, Berkeley, California 94720 USA}
\author{G.~Gabella}
    \affiliation{Department of Nuclear Engineering, University of California, Berkeley, California 94720 USA}
\author{J.~Gordon}
    \affiliation{Department of Nuclear Engineering, University of California, Berkeley, California 94720 USA}
\author{E.~Brubaker}
    \affiliation{Sandia National Laboratories, Livermore, California 94550 USA}

\date{\today}
\begin{abstract}
\begin{description}
\item[Background] Organic scintillators are widely used for neutron detection in both basic nuclear physics and applications. While the proton light yield of organic scintillators has been extensively studied, measurements of the light yield from neutron interactions with carbon nuclei are scarce.\item[Purpose] Demonstrate a new approach for the simultaneous measurement of the proton and carbon light yield of organic scintillators. Provide new carbon light yield data for the EJ-309 liquid and EJ-204 plastic organic scintillators. \item[Method] A 33~MeV $^{2}$H$^{+}$ beam from the 88-Inch Cyclotron at Lawrence Berkeley National Laboratory was impinged upon a 3-mm-thick Be target to produce a high-flux, broad-spectrum neutron beam. The double time-of-flight technique was extended to simultaneously measure the proton and carbon light yield of the organic scintillators, wherein the light output associated with the recoil particle was determined using $np$ and $n$C elastic scattering kinematics. \item[Results] The proton and carbon light yield relations of the EJ-309 liquid and EJ-204 plastic organic scintillators were measured over a recoil energy range of approximately 0.3 to 1~MeV and 2 to 5~MeV, respectively for EJ-309, and 0.2 to 0.5~MeV and 1 to 4~MeV, respectively for EJ-204. \item[Conclusions] These data provide new insight into the ionization quenching effect in organic scintillators and key input for simulation of the response of organic scintillators for both basic science and a broad range of applications. 
\end{description}

\end{abstract}

\maketitle

\section{Introduction}

An understanding of the response of organic scintillators to recoil nuclei is important in a wide range of basic and applied physics. In the search for dark matter, nuclear recoils in organic scintillators provide a means to detect hypothetical weakly and strongly interacting massive particles~\cite{Sekiya2003, Awe2018}. For neutrino studies, many large-scale (e.g., KamLAND~\cite{Eguchi2003}, SNO+~\cite{Andringa2016}, Borexino~\cite{Borexino2020}) and benchtop measurements (e.g., CHESS \cite{Caravaca2017}, FlatDot~\cite{Gruszko2019}, and MiniCHANDLER~\cite{Haghighat2020}) employ organic scintillators as the detection medium, wherein recoiling alpha particles from radioactive contaminants or protons from fast neutron interactions represent important sources of background~\cite{Westerdale2017}. Organic scintillators are also useful for neutron detection in a variety of national security and proliferation detection applications, and the proton light yield represents an important quantity in this regard. For example, the Single-Volume Scatter Camera---a compact scintillating medium designed for fast neutron source search, localization, and imaging---relies upon double-scatter kinematic reconstruction of neutron interactions, where the proton light yield is required to convert the measured scintillation light into proton recoil energy~\cite{Weinfurther2018, Manfredi2020proc}. In both basic science and applications, accurate modeling of the response of organic scintillators requires an understanding of the light yield of recoil particles. 

While the light output response of organic scintillators to recoil nuclei has been extensively studied, there remains significant disagreement in the literature as to the relationship between the specific luminescence and characteristics of the recoil particle~\cite{Yoshida2010,vonKrosigk2016,Christensen2018}. The extent of ionization quenching depends upon the ionization and excitation density in the scintillating medium, which increases with the stopping power of the recoil particle. As a result, the relationship between particle energy and scintillation light is nonlinear, and the amount of quenching increases with particle charge for a given recoil energy. The canonical Birks Law~\cite{Birks1951} (or generalizations thereof \cite{Chou1952, Wright1953, Hong2002, Yoshida2010}) are often used to describe the specific luminescence, but these models rely upon the knowledge of empirically-derived parameters. Currently, no theoretical formalism exists to accurately predict the ionization quenching effect in organic scintillators. Additional data are needed both to inform theoretical models of scintillator luminescence and characterize the response of commercial organic scintillators commonly used for basic nuclear physics and applications.  

In this work, the proton and carbon light yield of the EJ-309 and EJ-204 organic scintillators were measured simultaneously over a range of recoil energies. The EJ-309 liquid scintillator has pulse shape discrimination (PSD) properties~\cite{EJ309} and was selected as a standard of reference given the relatively large body of work characterizing the proton light yield of this medium~\cite{Takada2011,Enqvist2013,Pino2014,Lawrence2014,Tomanin2014,Iwanoska2015,Bai2017,BrownThesis, Brown2018,FREGEAU2019162301,Laplace2020_EJ309,Iwanoska2015}. The EJ-204 plastic scintillator has a rapid temporal response and long attenuation length~\cite{EJ204}, and it has been identified as a prime candidate for use in cutting-edge compact neutron imaging systems~\cite{Sweany2019}. In Sec.~\ref{exp}, an overview of the experimental setup and electronics configuration is provided. Section~\ref{analysis} consists of a description of the analysis methods used to obtain the proton and carbon light yield relations, including recoil energy determination, light output calibration, data reduction, and uncertainty quantification. In Sec.~\ref{results_and_discuss}, the measured EJ-309 and EJ-204 proton and carbon light yields are presented and compared to prior works where available. A discussion of the quenching factor, a metric commonly used to characterize the response of recoil nuclei in organic scintillators, is also provided. Concluding remarks are given in Sec.~\ref{summary}.

\section{Experimental Methods}
\label{exp}

The double time-of-flight (TOF) method of Brown et al.\ was extended to measure both the proton and carbon light yield of organic scintillators~\cite{BrownThesis,Brown2018}. Proton and carbon recoil energies were inferred via $np$ and $n$C elastic scattering kinematics allowing for event-by-event discrimination between these channels. Each nuclear recoil was associated with the scintillator light output to provide a continuous, simultaneous determination of both light yield relations. Measurements were performed for the EJ-309 liquid scintillator and the EJ-204 plastic scintillator, both from Eljen Technologies \cite{EJ309, EJ204}. 

\begin{figure}
\center
\includegraphics[width=0.47\textwidth]{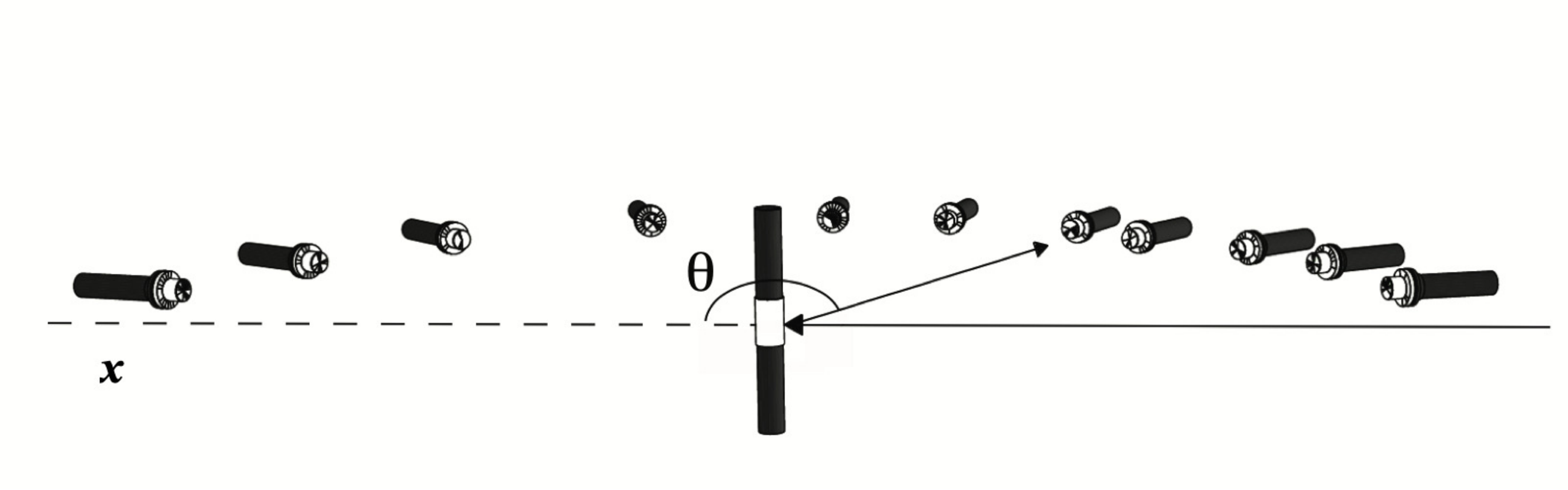}
\caption{Experimental setup for the EJ-204 proton and carbon light yield measurements. The neutron beam traveled from right-to-left along the $x$-axis through the dual-PMT target scintillator. Eleven observation detectors were positioned at $15-165^\circ$ with respect to the incoming neutron beam and at a distance of $1.5-1.6$~m from the target scintillator. The EJ-309 measurements employed a similar configuration. 
\label{exp-setup}}
\end{figure}

A broad spectrum neutron source was produced by impinging a 33~MeV $^{2}$H$^{+}$ beam from the 88-Inch Cyclotron at Lawrence Berkeley National Laboratory onto a 3-mm-thick Be target in the cyclotron vault~\cite{Harrig2018}. Neutrons were collimated using approximately 1~m of steel and 1.5~m of concrete and sand bags to yield a 20-cm-diameter open-air beam in the experimental area. The target scintillator (either EJ-309 or EJ-204) was placed in the neutron beam 7.5~m from the Be target and surrounded by eleven EJ-309 observation scintillators at angles spanning $15-165^\circ$ with respect to the incoming neutrons. The experimental setup is shown in Fig.~\ref{exp-setup} for the EJ-204 scintillator. In the EJ-309 measurement, the target scintillator was oriented horizontally and rotated approximately $60^\circ$ with respect to the incoming beam. Each observation detector was located approximately $1.5-1.6$~m from the target. The PSD capability of the observation scintillators allowed for selection of scattered neutrons, reducing false coincidences between neutron interactions in the target scintillator and background events in the observation cells. By measuring the neutron scattering angle, the neutron TOF between the target scintillator and the observation detector, and the neutron TOF between the target scintillator and the Be breakup target, neutron elastic scattering events on proton and carbon nuclei were distinguished using the known reaction kinematics.  

Each target scintillator was optically coupled to two gain-matched photomultiplier tubes (PMTs) using EJ-550 silicone grease. The EJ-309 target scintillator was a 5.08~cm~dia.~x~5.08~cm~len.\ right circular cylindrical liquid cell in a dual-window aluminum housing internally coated with EJ-520 titanium dioxide reflective paint~\cite{Eljen520}. The cell was coupled on each end to Hamamatsu H13795-100~PMTs biased at $-1750$~V and $-1800$~V. The EJ-204 target scintillator was a 5.08~cm~dia.~x~5.08~cm~len.\ right circular cylinder coupled on each base to a Hamamatsu H1949-51~PMT (biased at $-2350$~V and $-2280$~V for the upper and lower PMT, respectively). Each EJ-309 observation detector was a right cylindrical cell (5.08~cm~dia.~x~5.08~cm~len.)\ of thin aluminum housing coupled to a Hamamatsu PMT (either Type 1949-50 or 1949-51) via a borosilicate glass window and EJ-550 silicone grease. 

All PMTs were negatively biased using either a CAEN R1470ETD or CAEN NDT1470 power supply. Each PMT signal (as well as the cyclotron RF control signal) was connected to a CAEN V1730 500-MS/s, 14-bit digitizer. Data were acquired over a period of 18~h and 20~h with a beam current of approximately 12~nA for the EJ-309 and EJ-204 target scintillators, respectively. Full waveforms with global timestamps were recorded, with a 1.6~$\mu$s and 800~ns acquisition window for the signals from the target scintillator and observation detectors, respectively. The data acquisition system triggered on a coincidence between the two target PMTs (allowing for the rejection of uncorrelated dark current) and one of the observation detectors within a 400~ns coincidence window. The scintillator signal timing was determined using the CAEN digital constant fraction discrimination algorithm, with a 75\% fraction and a 4~ns delay. The timing pickoff for the cyclotron RF signal was determined using leading-edge discrimination. 

\section{Data Analysis}
\label{analysis}

The collected waveforms were processed using a custom object-oriented analysis suite leveraging components of the ROOT data analysis framework~\cite{ROOT}. Each waveform was baseline-subtracted before being integrated to provide a measure proportional to the total scintillation light \cite{Laplace2020_PH}. Waveform integration lengths of 400~ns and 200~ns were used for the EJ-309 and EJ-204 target scintillators, respectively, to ensure collection of at least 95\% of the scintillation light within the full acquisition window. For the dual-mounted target scintillators, the geometric mean of the individual pulse integrals from the two PMTs provided a measure of the total light output independent of interaction position for each event \cite{KnollRadiation2010}. The linearity of the response of the target PMTs was evaluated using a custom pulsed-LED circuit in accordance with the method of Friend et al.~\cite{Friend2011}, and the PMT response was determined to be linear within $\leq0.6\%$. 

\subsection{Recoil energy calculation}
\label{subsec:energy}

The incoming and outgoing TOF were calibrated using photons arising from deuterons impinging on the Be target. For the incoming TOF, a histogram of time differences between the cyclotron RF pulse and signals in the target scintillator yielded a distinct feature corresponding to photon events. The time calibration constant was then calculated using the speed of light and the flight path between the Be target and the target scintillator. The outgoing TOF calibrations were determined analogously using time differences between $\gamma$-ray interactions in the target scintillator and each observation scintillator. This procedure is described in detail in Ref.~\cite{Brown2018}.

\begin{figure}
	\center
	\includegraphics[width=0.47\textwidth]{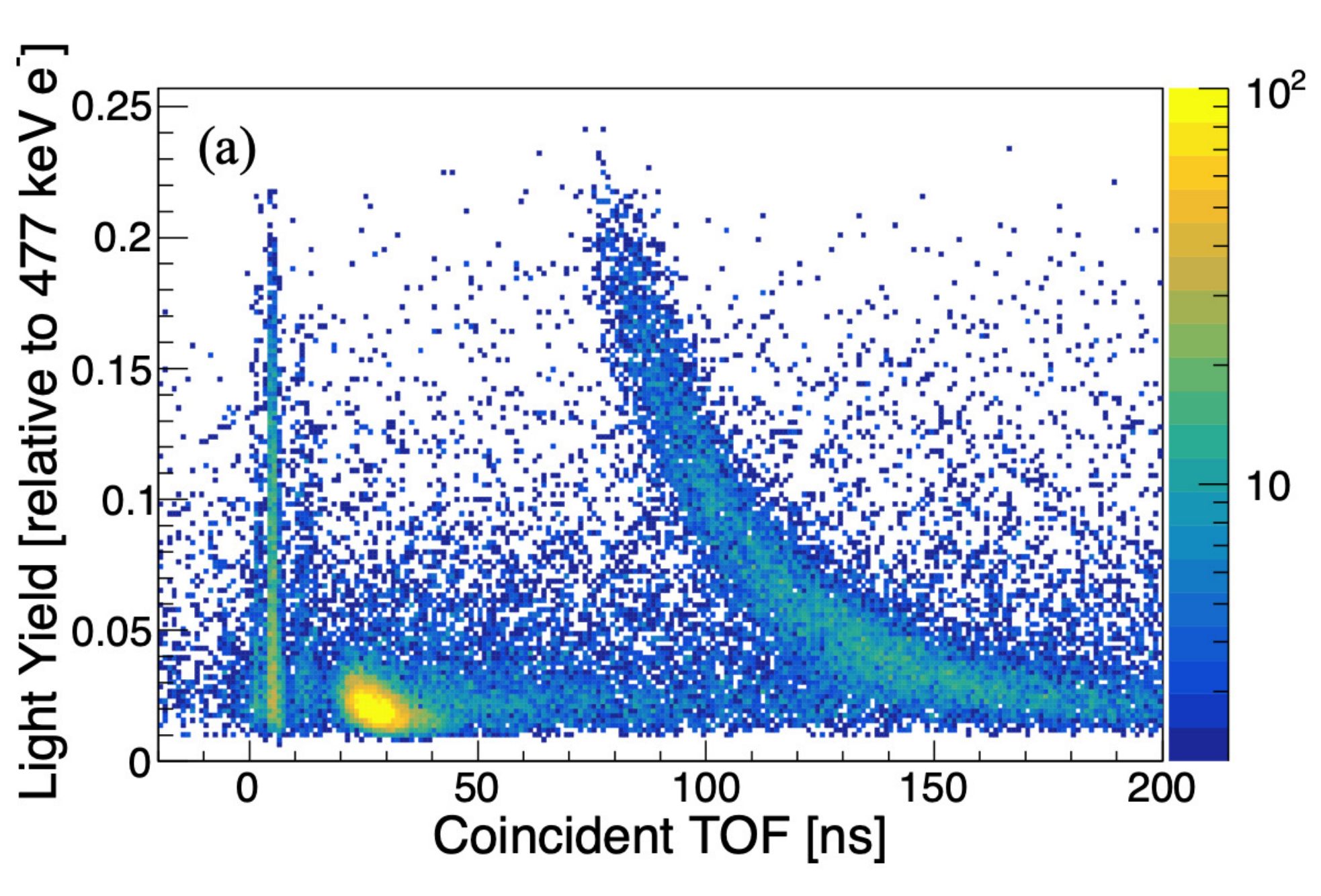}
	\includegraphics[width=0.47\textwidth]{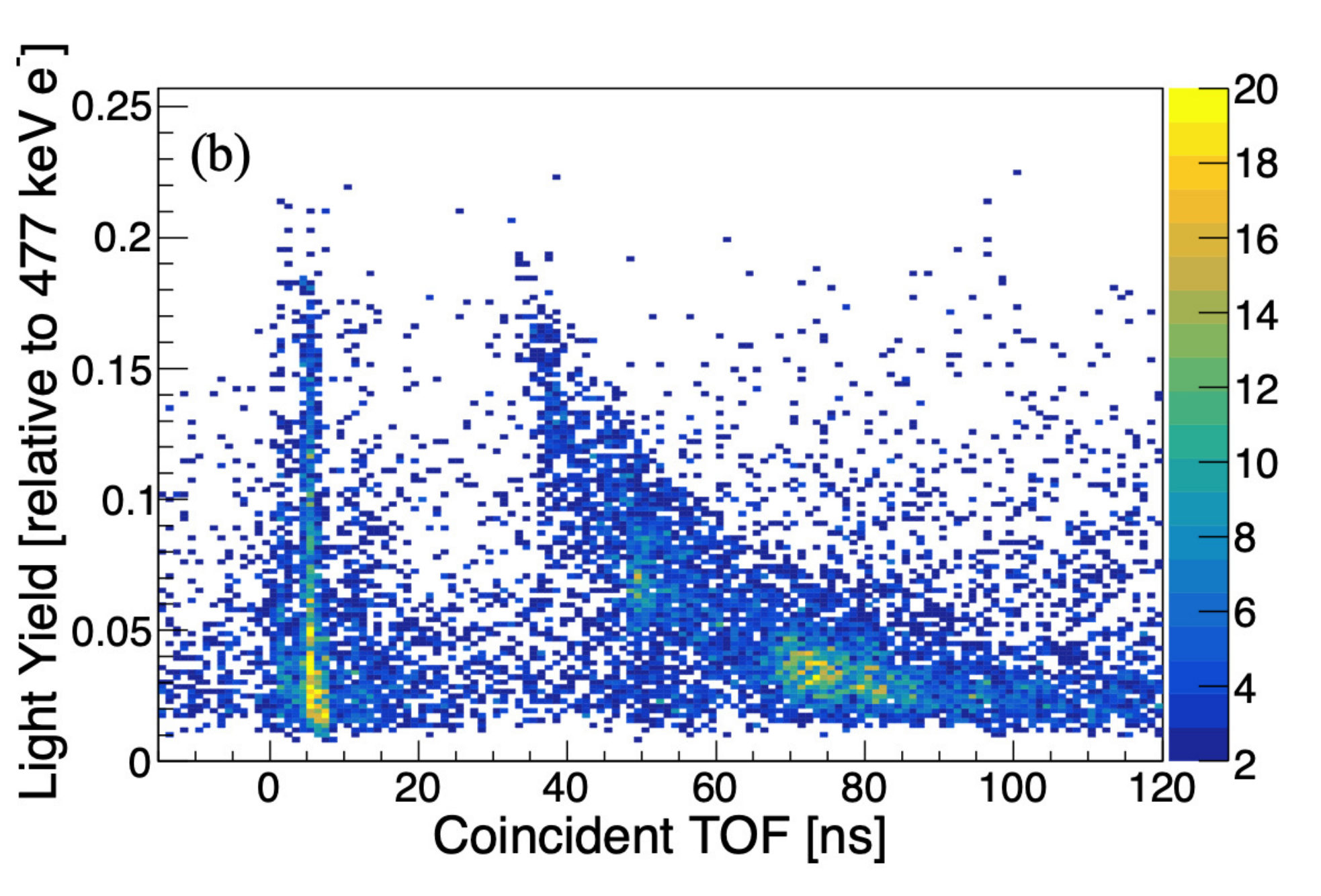}
	\caption{Light output of the EJ-204 target scintillator as a function of neutron TOF between the target and an observation detector located at (a) $30^\circ$ and (b) $150^\circ$ with respect to the incoming beam. \label{PhVsCoinTOF}}
\end{figure}

The outgoing neutron TOF was used to distinguish between proton and carbon recoils. Figure \ref{PhVsCoinTOF} shows the light yield of the EJ-204 target scintillator as a function of the scattered neutron TOF for two different observation detectors: one at 30$^\circ$ and the other at 150$^\circ$ with respect to the incoming beam. For the forward angle detector, three features are present: a narrow distribution centered at around 5~ns corresponding to ($\gamma$,$\gamma$) scatters, a low light feature from approximately 20 to 35~ns corresponding to $n$C elastic scattering events, and a large band from 80 to 200~ns corresponding to $np$ elastic scattering events. For neutrons scattered at $150^\circ$ with respect to the incoming beam, $np$ elastic scattering is kinematically forbidden. Only features from ($\gamma$,$\gamma$) and $n$C scattering events are present. 
Coincident TOF and target pulse height gates were established for each observation detector, enabling recoil particle identification on an event-by-event basis.

As the scattering angle, incoming TOF, and outgoing TOF are known for a given event, the system is kinematically overconstrained. For each neutron scattering event (identified using the PSD capability of the EJ-309 observation scintillators), the incoming TOF was calculated using the measured outgoing TOF, scattering angle, and known kinematics assuming both $np$ and $n$C elastic scattering reactions. Events were discarded in which the calculated incoming TOF differed from the measured incoming TOF by more than 10$\%$ of the 111.1~ns cyclotron period. This constraint removed potential ambiguity from frame overlap of the pulsed neutron beam as well as contributions from multiple scattering events in the target scintillator.

The proton recoil energy, $E_p$, was calculated using the known neutron scattering angle, $\theta$, for a given observation detector and the incoming neutron energy, $E_n$, as determined using the time difference between events in the target scintillator and the cyclotron RF signal:
\begin{equation}
\label{eq:proton}
E_p = E_n \sin^2\!{\theta}. 
\end{equation}
Similarly, nonrelativistic scattering kinematics provided the carbon recoil energy:
\begin{equation}
\label{eq:carbon}
E_\text{C} = E_n\left[1 - \left(\frac{\cos{\theta} + \sqrt{A^2 - \sin^2\!{\theta}}}{A+1}\right)^2\right],
\end{equation}
where $A = 12$ is the mass number of carbon.\footnote{As the scintillating media were not isotopically enriched, the carbon recoil energy calculation neglects the stable $^{13}$C isotope at 1.1\% abundance.} The recoil energy uncertainty was determined via a Geant4 simulation of the experimental setup~\cite{Geant4}. For each simulated event, the recoil energy reconstructed with Eqs.~(\ref{eq:proton}) and (\ref{eq:carbon}) was compared to the ground-truth simulated recoil energy. This yielded a measure of the energy-dependent ion energy resolution taking into account the angular spread of the scintillator detectors, the temporal uncertainty of the incident neutron TOF (which was dominated by the spatial spreading of the $^{2}$H$^{+}$ beam), and the uncertainty in the incoming neutron flight pathlength. These uncertainty relations were used to determine the recoil energy binning for the proton and carbon light yield data given below.

\subsection{Scintillator light output calibration} 
\label{LOcalib}

\begin{figure}
	\center
	\includegraphics[width=0.47\textwidth]{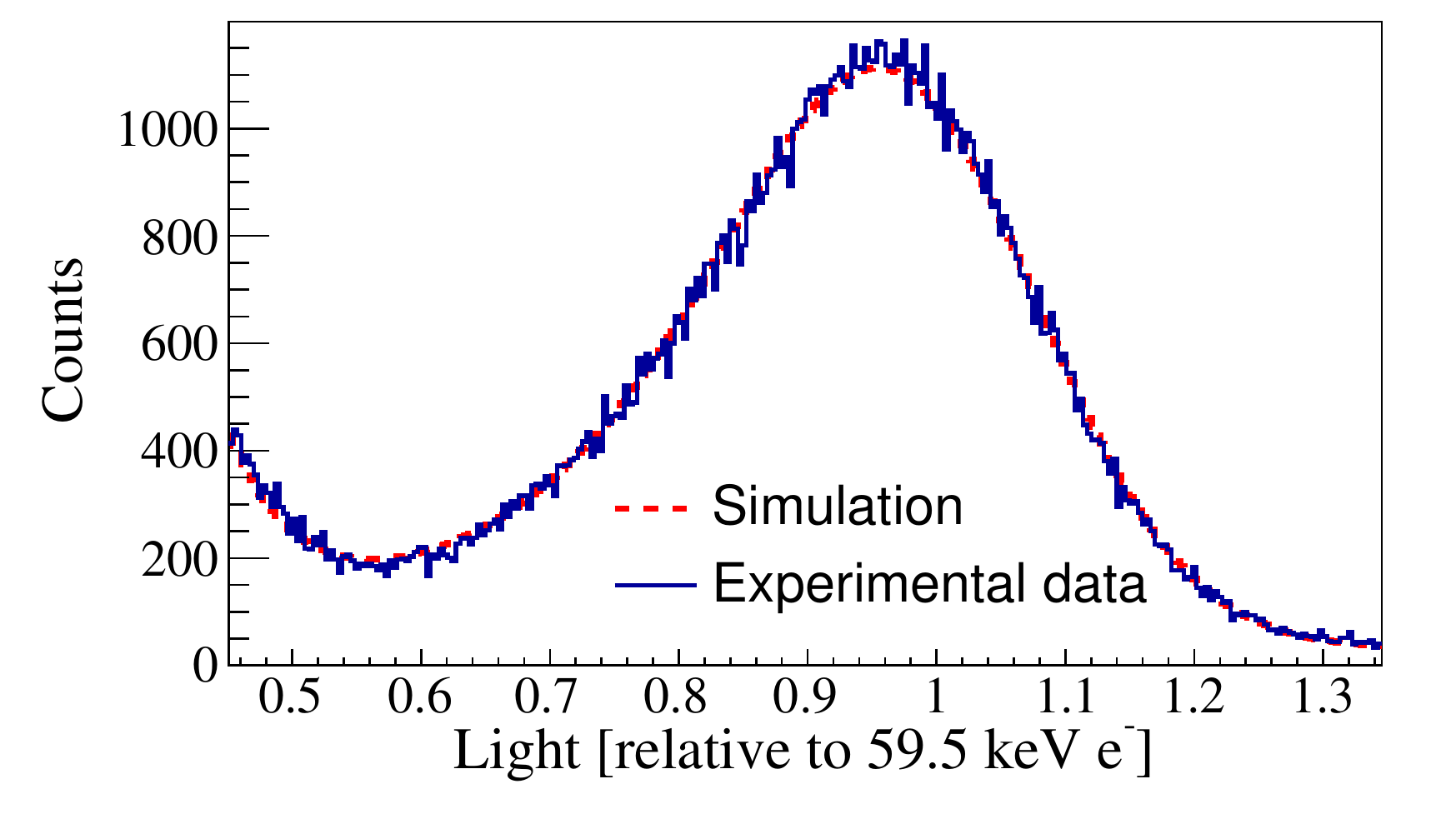}
	\caption{Light output spectrum (blue histogram) produced by 59.5~keV $\gamma$ rays incident on the EJ-309 target scintillator. The dashed red curve represents a simulated light output spectrum folded with a resolution function. \label{lightCalibration}}
\end{figure}

The scintillator light output was calibrated using the 59.5~keV $\gamma$ ray from an $^{241}$Am source positioned approximately 5~cm from the target scintillator, as described in Ref.~\cite{Gabella2020Placehold}. A representative $^{241}$Am light output calibration spectrum is shown for EJ-309 in Fig.~\ref{lightCalibration}. A Geant4 simulation was used to model the energy deposition spectra of recoil electrons arising from monoenergetic 59.5~keV $\gamma$ rays incident on the target scintillators. The model geometry also considered the surrounding materials, including the PMTs, mounts, and scintillator housing. The energy deposited by each simulated particle track was translated to light output taking into account the electron light nonproportionality, as further discussed below. The accumulated light output spectra were then convolved with a detector resolution function~\cite{Dietze1982}, and the difference between the resulting distribution and the measured spectrum was minimized while varying the calibration parameter (i.e., the scaling factor between the raw integrated channel and light output) and the three parameters of the resolution function. This minimization was executed using the SIMPLEX and MIGRAD algorithms from the ROOT Minuit2 package \cite{ROOT}. 

Given the lack of EJ-204 and EJ-309 electron light nonproportionality measurements in the literature, data on alternate media were used to convert the simulated energy deposition spectra to light output. In particular, the EJ-200 nonproportionality curve from Payne et al.~\cite{Payne2011} was applied to generate the modeled EJ-204 light output spectrum. As EJ-200 and EJ-204 both employ a polyvinyltoluene (PVT) solvent, it is reasonable to assume that these media will have very similar electron light nonproportionality curves given that ionization quenching is a primary process~\cite{BirksTheoryAndPractice1964}. To generate the modeled EJ-309 light output spectrum, two electron light nonproportionality curves for similar media were implemented and their impact on the calibration parameters was evaluated. Specifically, the ``liquid scintillator'' and EJ-301 nonproportionality curves from Payne et al.~\cite{Payne2011} and Swiderski et al.~\cite{Swiderski2012}, respectively, were each applied anchoring the electron light nonproportionality at 59.5 keVee, and the calibration parameters obtained using the two relations agreed within uncertainties. That is, the $^{241}$Am calibration for the EJ-309 target was insensitive to the choice of electron light nonproportionality data used to generate the modeled spectrum. 

To facilitate comparison against previous EJ-309 and EJ-204 proton light yield measurements in which the light output calibration was performed using a $^{137}$Cs source~\cite{Laplace2020,Laplace2020_EJ309}, the light unit was defined such that the light output corresponding to the Compton edge of a 662~keV $\gamma$ ray was set to 1. This is in contrast to an electron-equivalent light unit, commonly employed to describe scintillator light output, which implicitly assumes the proportionality of the scintillator response to electron recoils. This is often an inappropriate assumption particularly at low recoil energies, including in the case of liquid and plastic organic scintillators~\cite{Nassalski2008,Payne2011}. For the in-beam measurement, the PMT bias voltages were set with single photon sensitivity to permit detection of the more strongly quenched carbon recoils. As such, the Compton edge of the 662~keV $\gamma$ ray was off scale given the 14-bit resolution and 2~V dynamic range of the digitizer. Cross-calibration datasets were taken for each scintillating medium in which the PMT bias voltages were set to allow the $^{241}$Am and $^{137}$Cs calibration points to appear together in the dynamic range. The fits proceeded as described above (for each source spectrum separately), and the ratio between the two points enabled a conversion from the relative light scale established by the light output resulting from a 59.5~keV photoelectron to that defined by a 477~keV electron (corresponding to the Compton edge of a 662~keV $\gamma$ ray). In the case of EJ-204, the ratio between these two points was consistent with the ratio provided by the EJ-200 electron light nonproportionality curve from Payne et al.~\cite{Payne2011}. For EJ-309, the measured ratio disagreed with that provided by the ``liquid scintillator''~\cite{Payne2011} and EJ-301~\cite{Swiderski2012} data by 4.3\% and 2.6\%, respectively.   

\begin{figure}
	\center
	\includegraphics[width=0.47\textwidth]{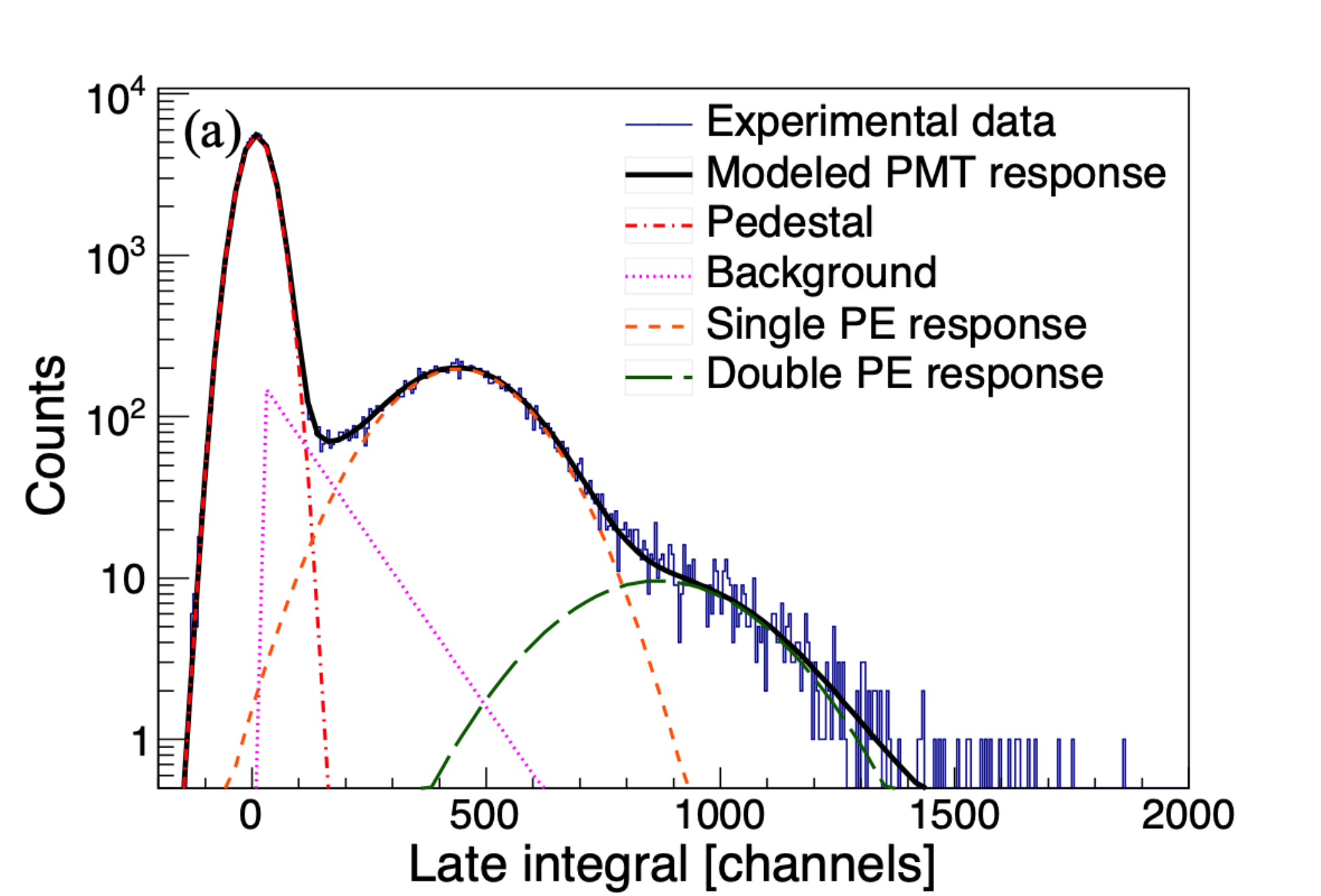}
	\includegraphics[width=0.47\textwidth]{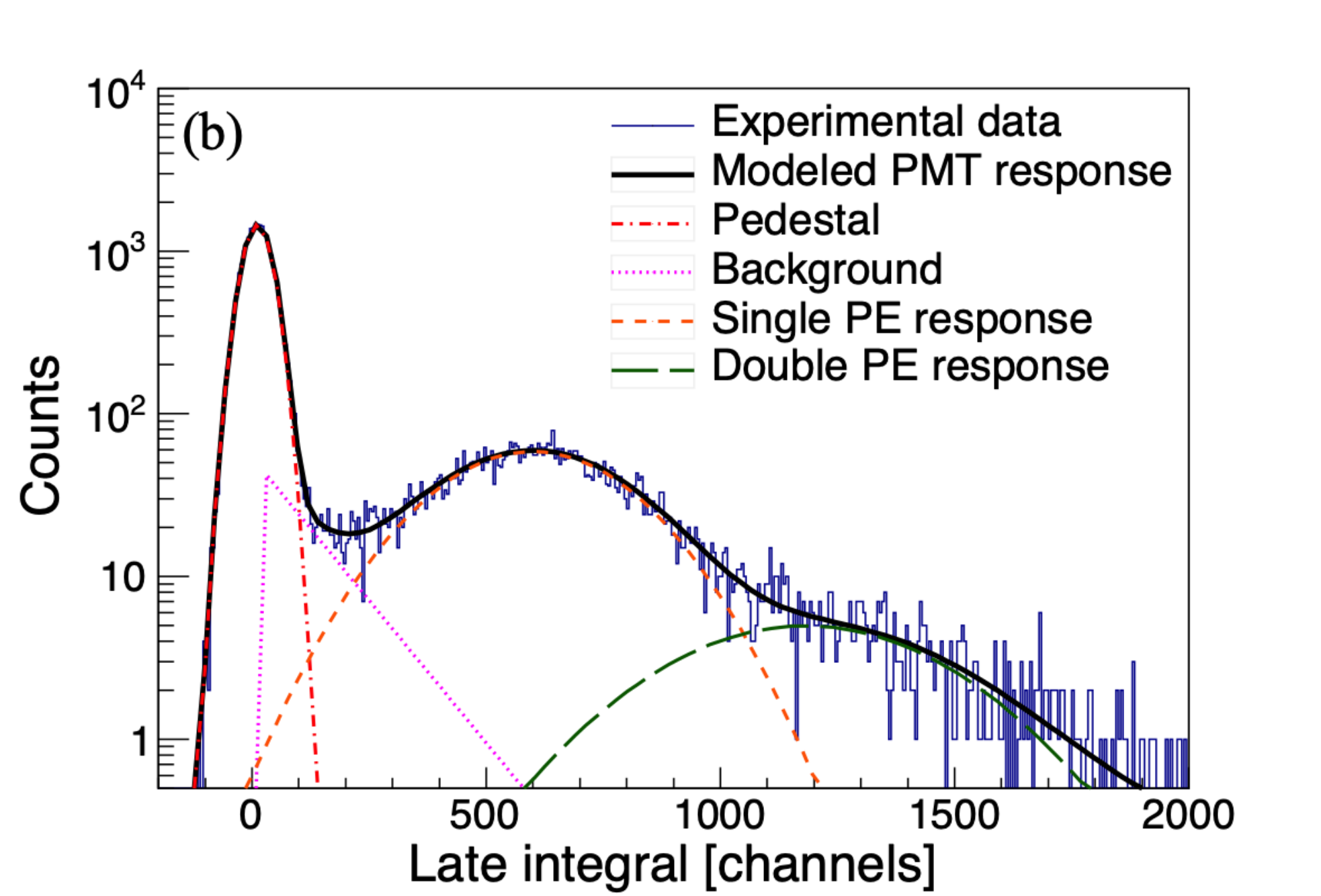}
	\caption{Low-amplitude light output distributions for the (a) calibration and (b) in-beam EJ-204 datasets, where a 100~ns integral was taken 500~ns following a scintillation pulse. The measured data (blue histogram) are fitted with a PMT response function (black curve) composed of the following individual components: single photoelectron response (orange dashed curve), double photoelectron response (green long dashed curve), pedestal (red dash-dotted curve), and an exponential background (dotted magenta curve). \label{singlepe}}
\end{figure}

For the EJ-204 experiment, the PMT response changed significantly between the $^{241}$Am calibration measurement and the in-beam neutron measurement. Such a rate-dependent gain variation has been previously observed for PMTs with a linear-focused dynode structure~\cite{Yamashita1978}. The magnitude of the effect was evaluated by comparing the response of a single photoelectron pulse in each of the target PMTs for the calibration and runtime settings. Figure~\ref{singlepe} shows late-time pulse integral distributions for the EJ-204 calibration and in-beam data, where a 100~ns integral was taken 500~ns following a scintillation pulse to emphasize single and few photoelectron events. A clear, discrete change in the photodetector gain is observed between the calibration and runtime settings. The average charge of a single photoelectron, which is proportional to the PMT gain coefficient, was extracted for each target PMT by fitting the measured distributions using the PMT response function from Bellamy et al.~\cite{Bellamy1994} and a binned-likelihood estimation. The ratio of the geometric average of the mean charge of the single photoelectron response of each target PMT for the in-beam to the calibration setting, determined as $1.40\pm0.03$, was then subsequently applied to correct for this effect. A comparable study of the EJ-309 data showed no statistically significant discrepancy in the PMT response for the calibration and runtime settings.

\subsection{Proton and carbon light yield determination}

\begin{figure}
	\center
	\includegraphics[width=0.47\textwidth]{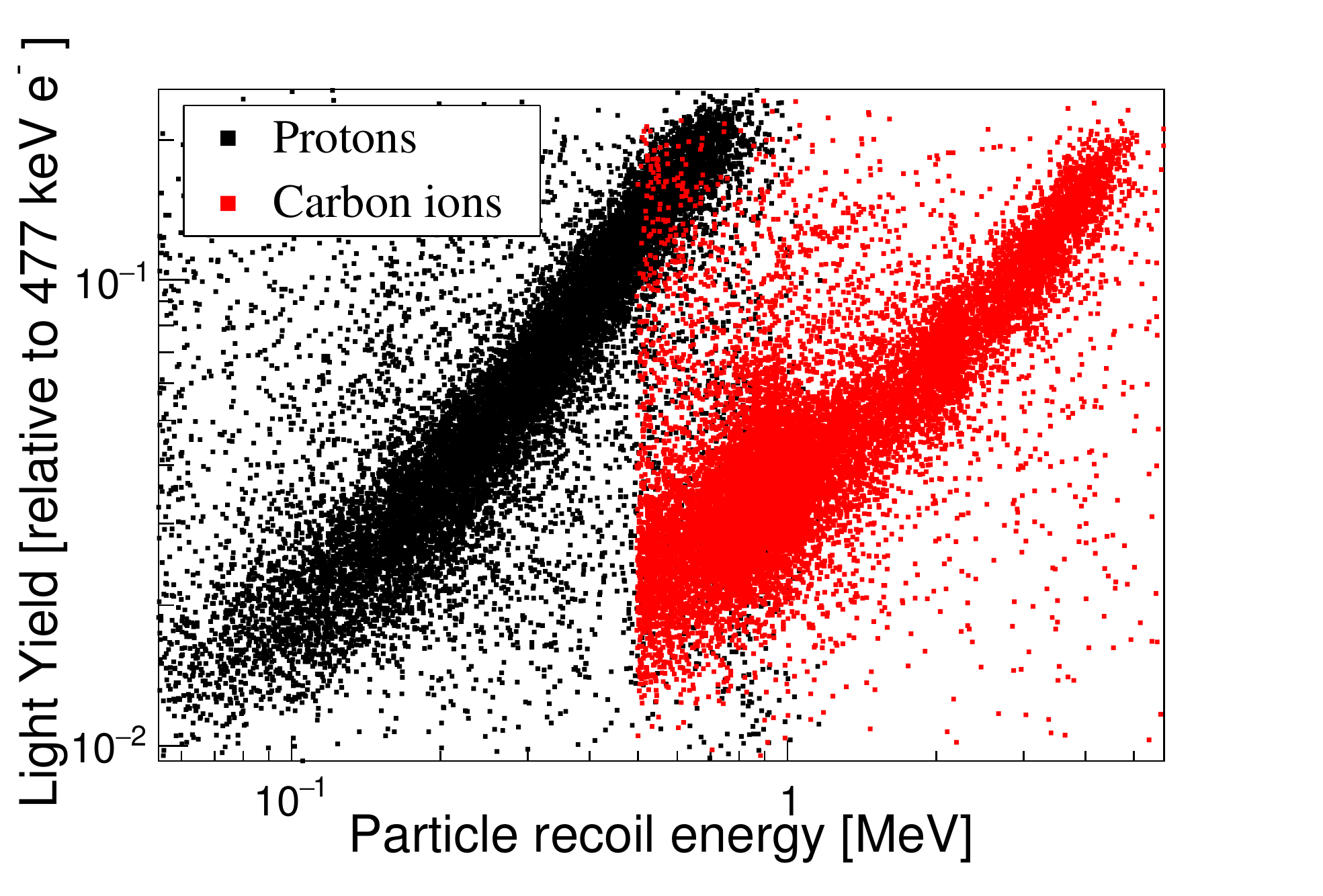}
	\caption{Light output of EJ-204 as a function of the particle recoil energy for proton (black) and carbon (red) recoils. A recoil energy threshold was applied to the carbon ions for visual clarity. \label{2dplots}}
\end{figure}

\begin{figure*}
\centering
	\subfloat[$\overline{\mathrm{E}_\mathrm{p}}$ = 223 keV. \label{lowest}]{\includegraphics[width=0.50\textwidth]{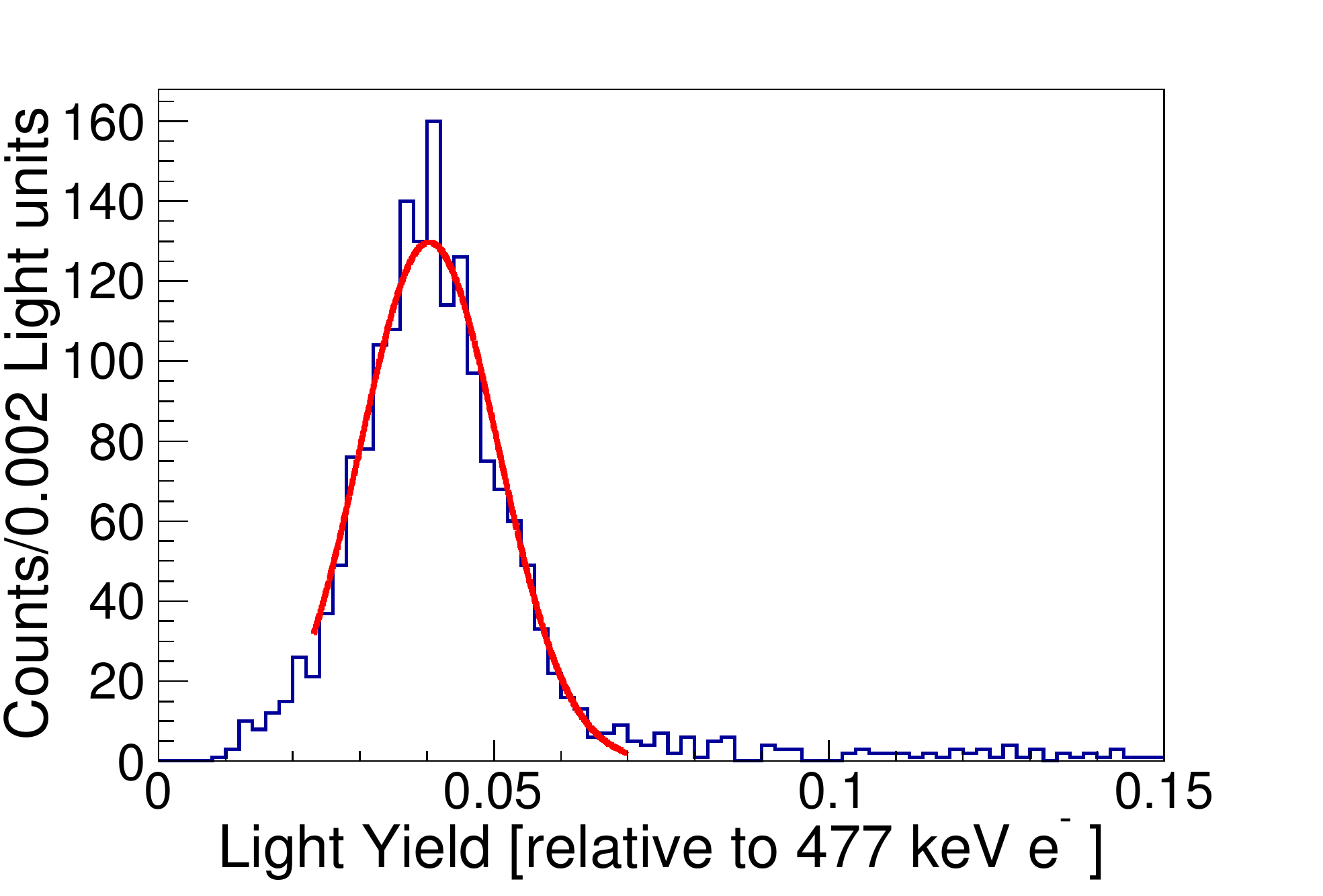}}
	\subfloat[$\overline{\mathrm{E}_\mathrm{p}}$ = 512 keV. \label{512}]{\includegraphics[width=0.50\textwidth]{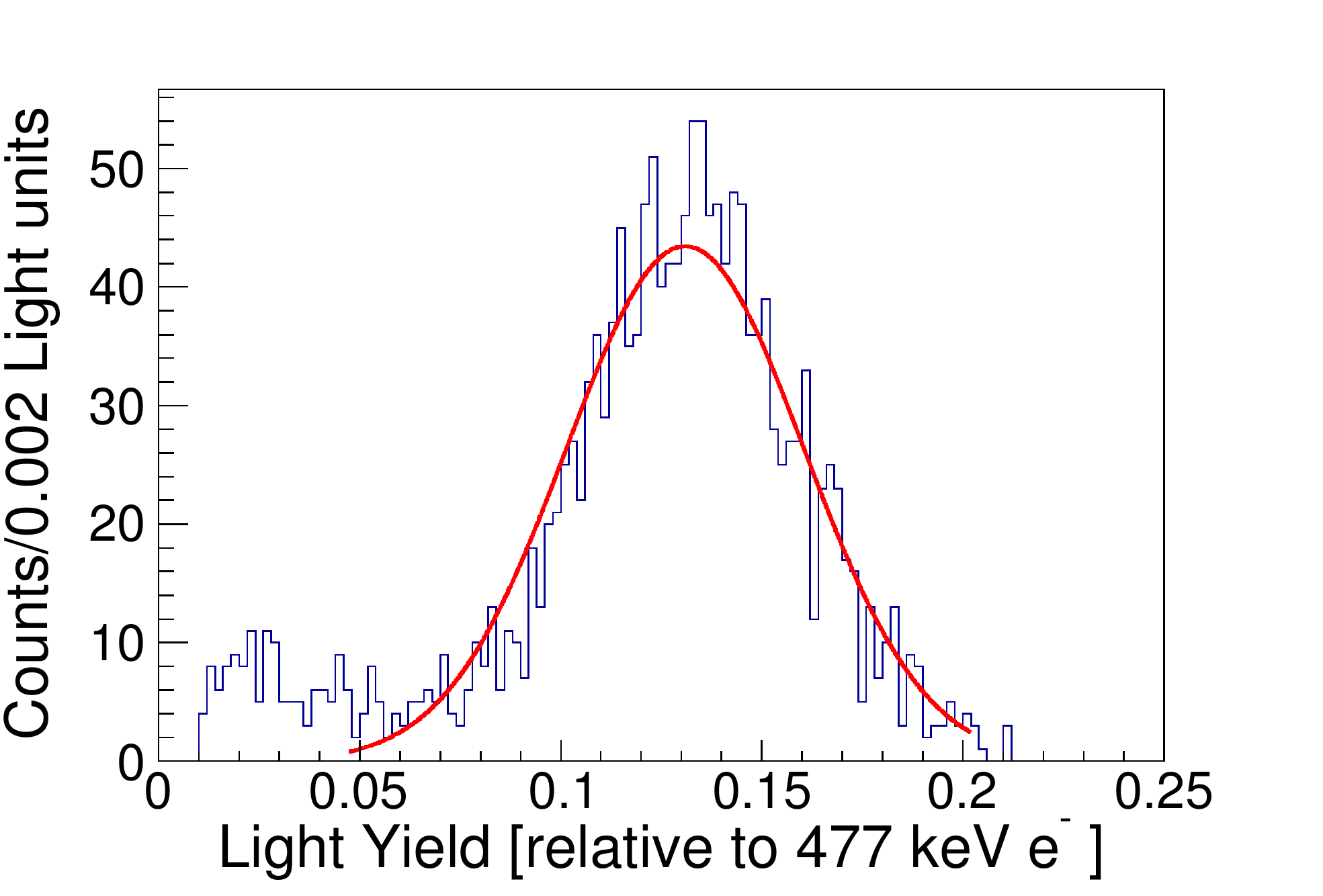}}
	\newline
	\subfloat[$\overline{\mathrm{E}_\mathrm{C}}$ = 1090 keV. \label{1090}]{\includegraphics[width=0.50\textwidth]{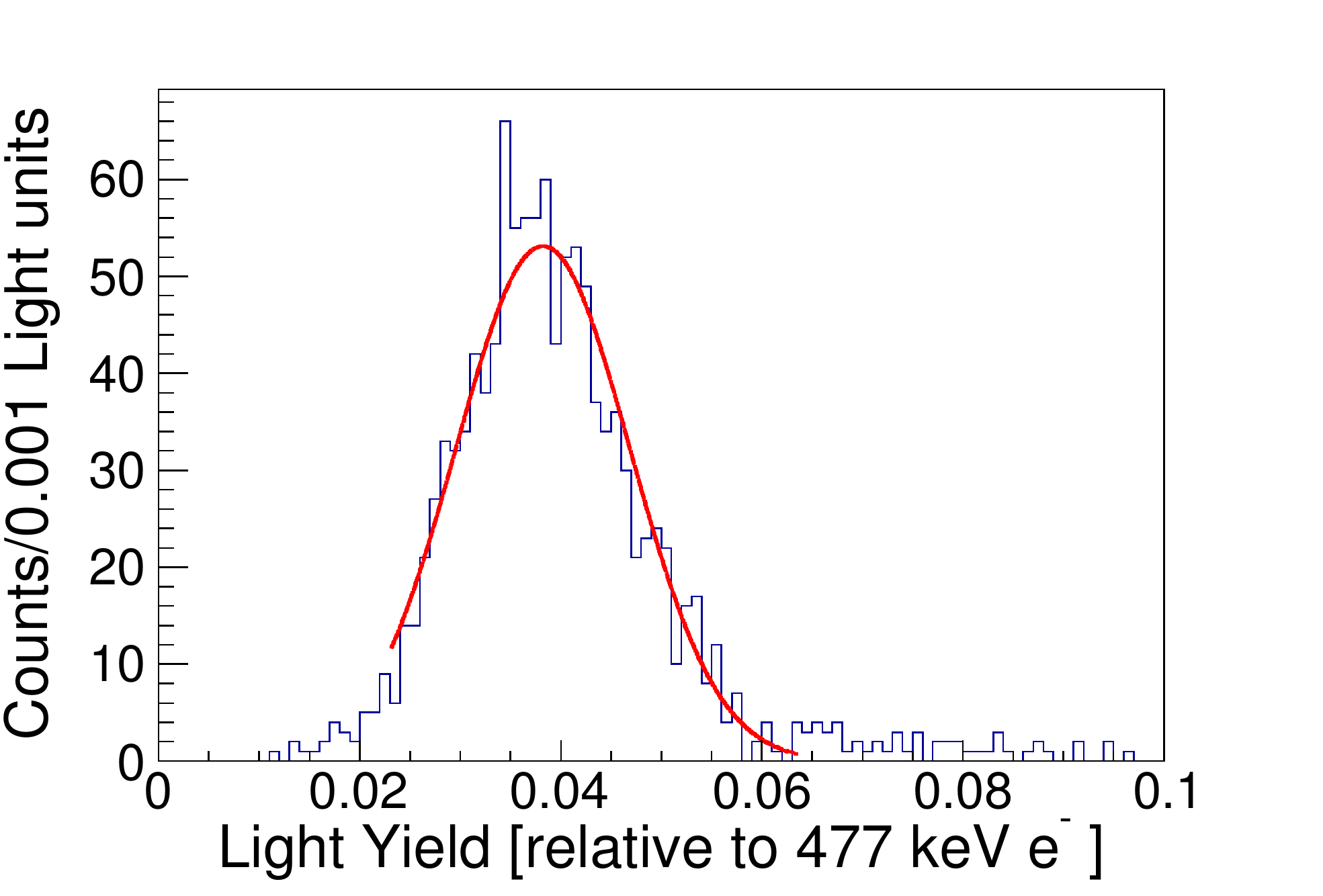}}
	\subfloat[$\overline{\mathrm{E}_\mathrm{C}}$ = 1374 keV. \label{1374}]{\includegraphics[width=0.50\textwidth]{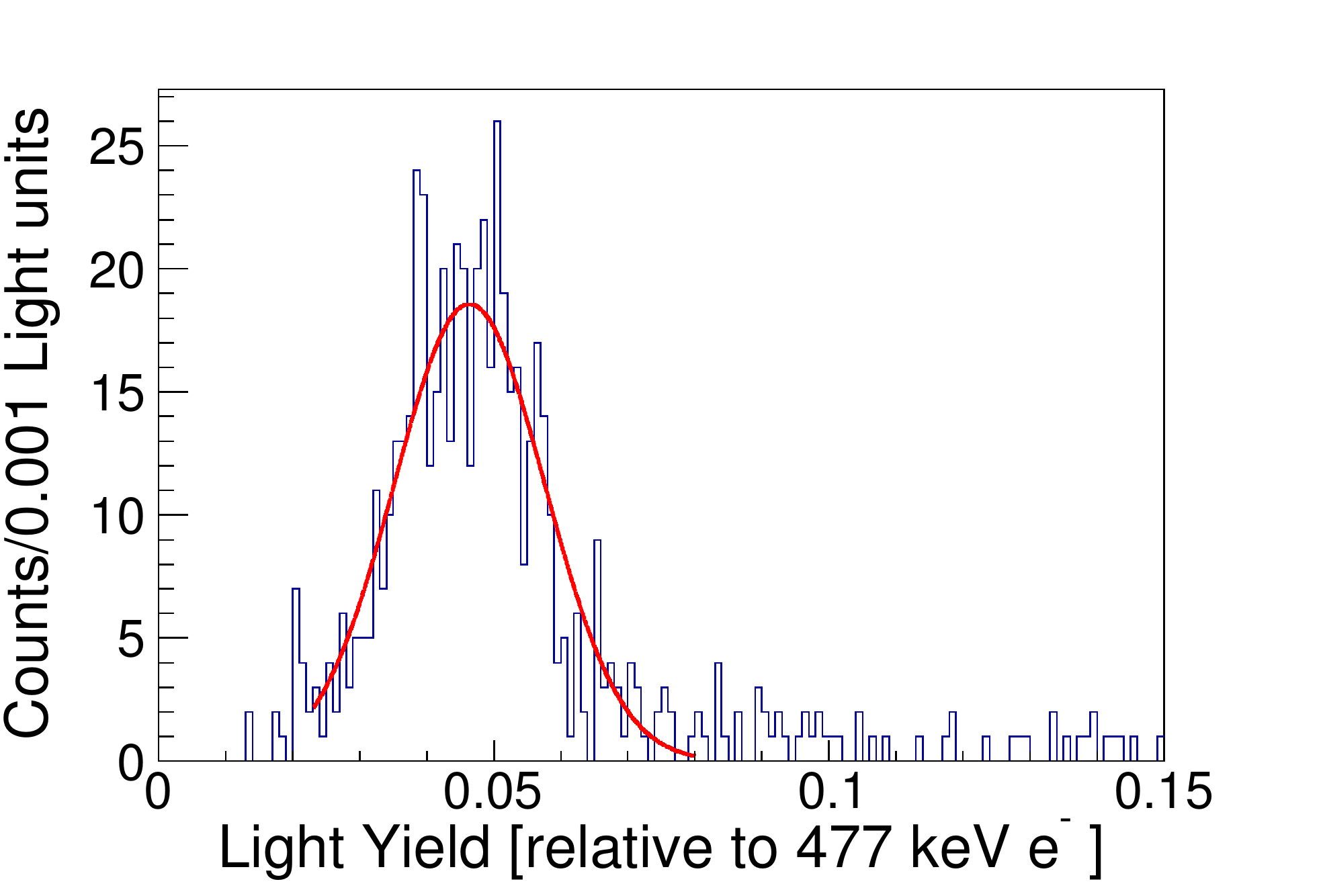}}
    \newline
	\subfloat[$\overline{\mathrm{E}_\mathrm{C}}$ = 2375 keV. \label{2375}]{\includegraphics[width=0.50\textwidth]{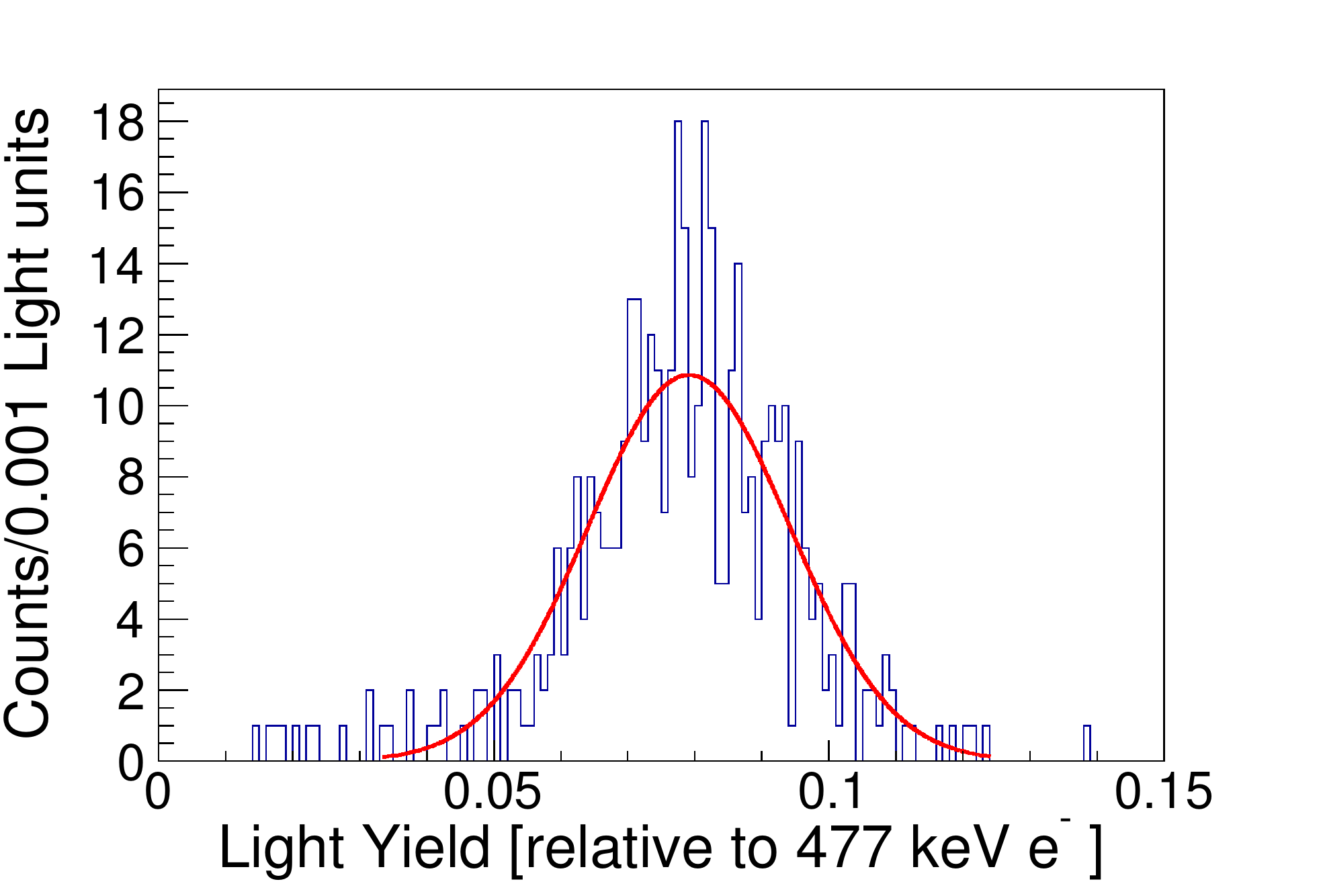}}	
	\subfloat[$\overline{\mathrm{E}_\mathrm{C}}$ = 4009 keV. \label{4009}]{\includegraphics[width=0.50\textwidth]{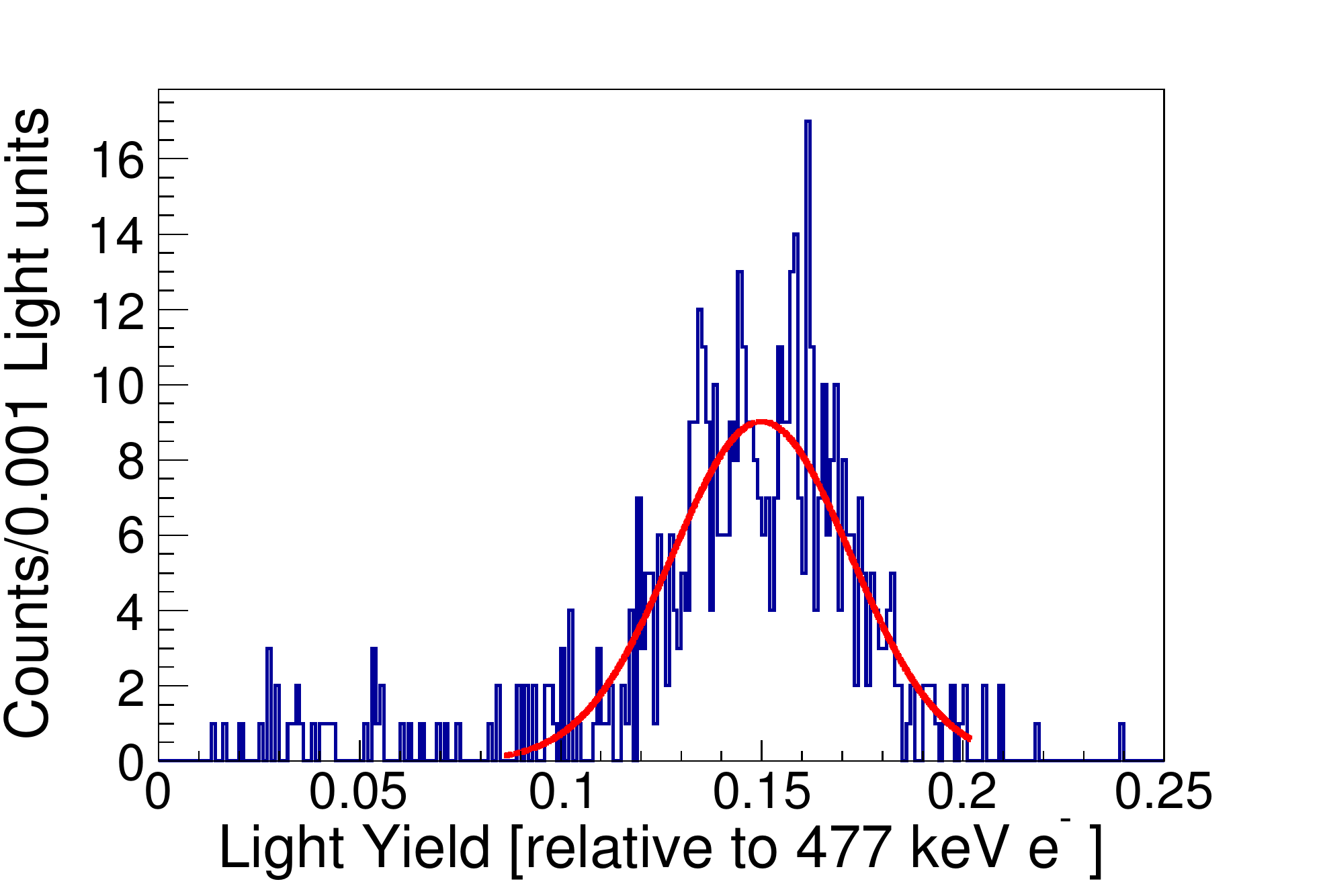}}	
	\newline
	\caption{(Color online) EJ-204 relative light output spectra (blue histograms) fit with a normal distribution (red curves). Subpanels (a) and (b) correspond to the minimum and maximum proton recoil energy bins with average energy $\overline{\mathrm{E}_\mathrm{p}}$. Subpanels (c) through (f) correspond to carbon recoil energy bins with average energy $\overline{\mathrm{E}_\mathrm{C}}$. The centroid of the normal distribution corresponds to the mean light production for the $np$ or $n$C elastic scattering events within the bin. \label{sliceFits}}
\end{figure*}

Figure~\ref{2dplots} shows a plot of the EJ-204 light output as a function of the reconstructed recoil energy for both protons and carbon ions. To reduce the data to the corresponding light yield relations, the recoil energy axis was binned according to the heteroskedastic energy resolution function described in Sec.~\ref{subsec:energy}. That is, the energy bin width varies as a function of energy reflective of the recoil
energy uncertainty. Each bin was projected onto the light output axis and fit with a Gaussian distribution using a binned-likelihood estimation to determine the light yield at the corresponding recoil energy. The fit range in each bin was set iteratively. First, the peak centroid and standard deviation were obtained, and then the feature was refit over a range centered about the centroid with a width of $\pm3\sigma$. Upper and lower range limits were further applied corresponding to effects arising in the dual-PMT target from signal saturation and the detection threshold, respectively.  Figure~\ref{sliceFits} provides several examples of these individual bin fits for both the EJ-204 proton and carbon light yield data. 

\subsection{Uncertainty quantification}

The sensitivity of the light yield relation to various analysis parameters was quantified using a Monte Carlo simulation in which each parameter was varied by sampling a normal distribution with the standard deviation given by the parameter uncertainty. These parameters included the TOF calibration constants, the measured detector locations, and the pathlength between the Be target and the terminus of the beam pipe in the experimental area, which served as the origin of the measurement coordinate system. Data reduction was then repeated with each parameter set to evaluate the effect on the light yield. Although each parameter only had a direct effect on the proton or carbon recoil energy, the fixed-bin structure of the energy axis led to changes in the composition of events (and therefore the light yield) in each bin. 

The systematic uncertainty in the single photoelectron response (determined in both calibration and runtime settings to assess potential gain shift) was taken as the quadrature sum of the standard deviation of the mean single photoelectron charge obtained for each target PMT for various start times in the tail of the scintillation pulse. For the calibration setting, the uncertainty in the single photoelectron response for EJ-309 and EJ-204 was determined to be 2.6\% and 1.7\%, respectively. For the runtime setting, the stability of the PMT gain was also assessed by partitioning the beam data into subsets and quantifying the single photoelectron response as a function of time. No drift was observed outside of the 4.9\% and 1.0\%  uncertainty of the single photoelectron response determined for the EJ-309 and EJ-204 scintillators, respectively. 

The $^{241}$Am light output calibration carried a 0.7\% uncertainty, which was obtained by evaluating the sensitivity of the centroid of the photopeak for 59.5~keV $\gamma$ rays to the ranges of the calibration fit and the propagation of errors from the cross-calibration. The uncertainty in the determination of the light output calibration, PMT gain instability, and rate-dependent gain shift were added in quadrature with the uncertainty obtained from the Monte Carlo simulation to provide the total uncertainty on the light yield relations.

\section{Results and Discussion}
\label{results_and_discuss}

\begin{figure}
	\center
	\includegraphics[width=0.49\textwidth]{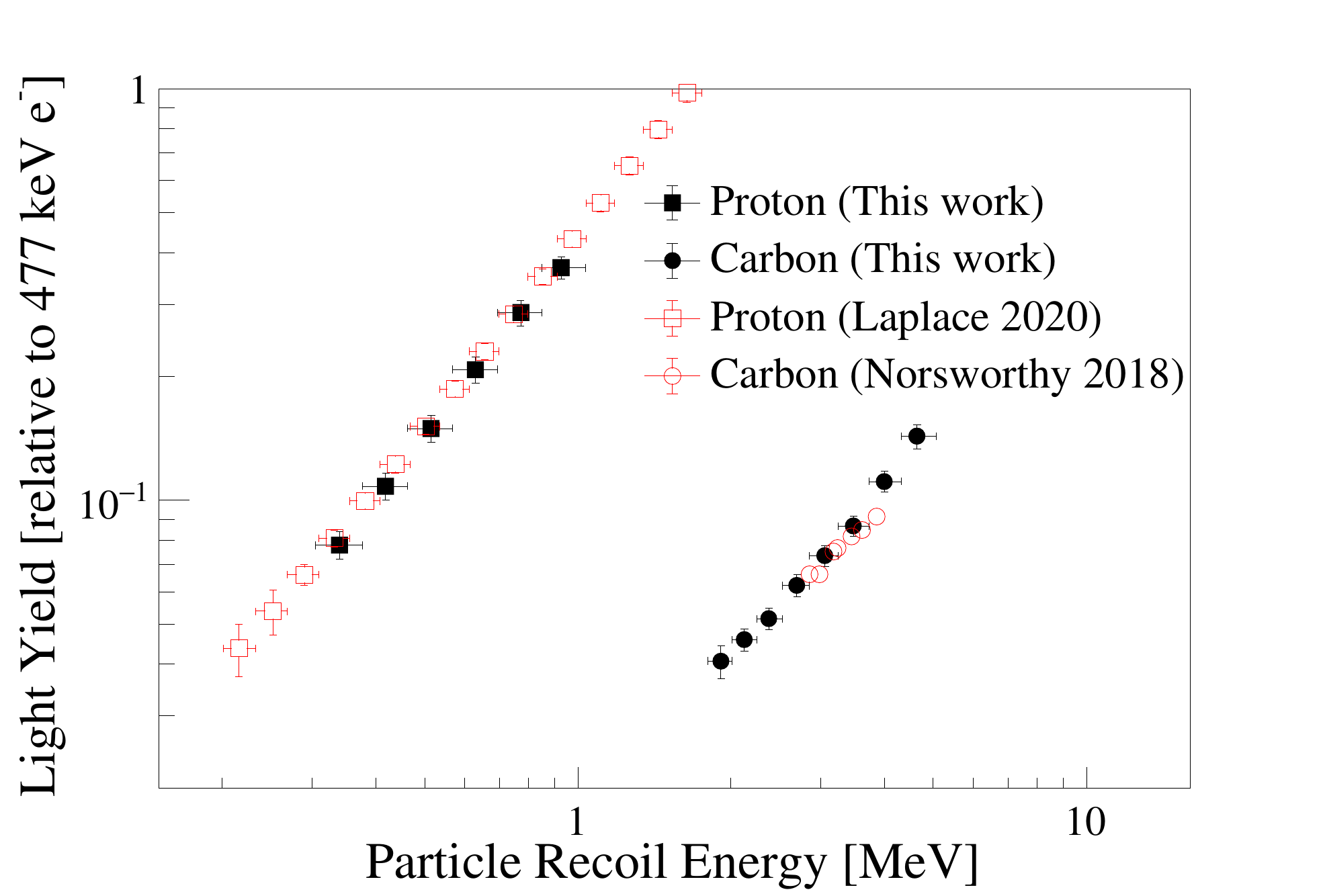}
	\caption{EJ-309 proton and carbon light yield relations. The square and circle symbols correspond to EJ-309 proton and carbon light yield data, respectively. The black filled symbols represent the present work. The red open symbols represent previous EJ-309 proton (Laplace 2020: ~\cite{Laplace2020_EJ309}) and carbon (Norsworthy 2018:~\cite{Norsworthy2018}) light yield measurements. The error bars on the abscissa represent a bin width and those on the ordinate axis include both the statistical and systematic uncertainties. \label{EJ309ResultsPlots}}
\end{figure}

Figure~\ref{EJ309ResultsPlots} displays the measured EJ-309 proton and carbon light yield relations alongside previous works. The EJ-309 proton light yield is compared to a double TOF measurement from Laplace et al.~\cite{Laplace2020_EJ309}, and the results are consistent within the estimated uncertainties. The EJ-309 carbon light yield data are compared to a measurement from Norsworthy et al.~\cite{Norsworthy2018}, which used a mono-energetic neutron source and an indirect technique to provide the carbon light yield relation in electron-equivalent light units. Norsworthy et al.\ calibrated the light output using the Compton edge of the 662~keV $\gamma$ ray from a $^{137}$Cs source~\cite{Norsworthy2018}. The light units were converted to that of the present work by setting 477~keVee equivalent to 1 light unit, and the measurements agree within 2$\sigma$. The consistency of the measured EJ-309 proton and carbon light yield relations with literature measurements lends confidence both to the extension of the double TOF technique and its application towards the generation of novel carbon light yield data.

\begin{figure}
	\center
	\includegraphics[width=0.49\textwidth]{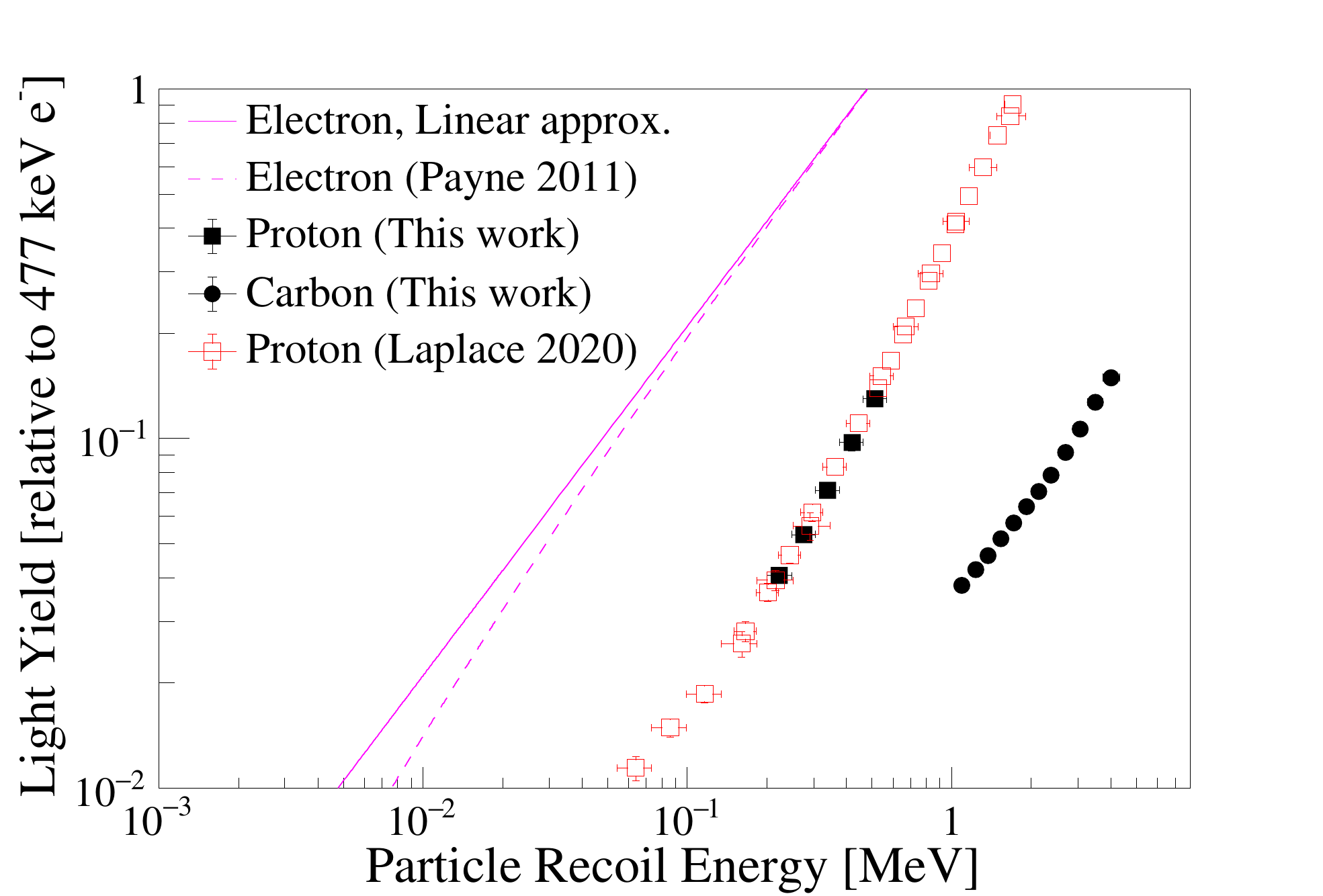}
	\caption{EJ-204 electron, proton, and carbon light yield relations. The square and circle symbols correspond to EJ-204 proton and carbon light yield data, respectively. The filled symbols represent the present work. The red open squares correspond to a previous EJ-204 proton light yield measurement (Laplace 2020:~\cite{Laplace2020}). The error bars on the abscissa represent a bin width and those on the ordinate axis include both the statistical and systematic uncertainties. The EJ-200 electron light yield (Payne 2011:~\cite{Payne2011}) is represented by the dashed magenta line. A linear approximation for the electron light yield is shown with the continuous magenta line for comparison. \label{EJ204ResultsPlots}}
\end{figure}

The measured EJ-204 proton and carbon light yield relations are shown in Fig.~\ref{EJ204ResultsPlots}. The EJ-204 proton light yield is compared to previous measurements from Laplace et al.~\cite{Laplace2020} that also used a double TOF technique. Again, the results are in agreement within the estimated uncertainties. No prior measurements of the EJ-204 carbon light yield exist. The measured EJ-309 and EJ-204 light yield data along with their associated uncertainties are tabulated in the appendix~\ref{LYDataPoints}.

The scintillation response to recoil nuclei is often characterized in terms of a quenching factor, defined as the ratio between the light observed from a nuclear recoil relative to an electron of the same energy~\cite{Yoshida2010,Reichart2012,Awe2018}. In the event that the scintillation light produced is proportional to the electron energy deposited in the medium, the quenching factor provides a means to quantify the reduction in light output due to ionization quenching. However, for electrons with relatively high stopping power, the electron light yield of organic scintillators is nonproportional to the energy deposited. For comparison, the EJ-204 electron light yield is also illustrated in Fig.~\ref{EJ204ResultsPlots}. The solid line corresponds to a linear approximation of the electron light yield anchored at 477~keV. The dashed line represents the EJ-200 electron light nonproportionality measurement from Payne et al.~\cite{Payne2011}, which deviates from the linear approximation due to the higher excitation and ionization density produced by slower electrons.\footnote{The EJ-200 and EJ-204 electron light nonproportionality are assumed to be equivalent as both media are comprised of a PVT base and ionization quenching reduces the primary excitation efficiency~\cite{BirksTheoryAndPractice1964}.} This deviation is manifest via the same mechanism---high ionization and excitation density---that results in the reduced scintillation efficiency observed for recoil nuclei. In this case, the quenching factor determined with respect to electrons no longer provides a measure of ionization quenching, as both the numerator and denominator vary due to the ionization quenching effect. It is worth noting that the electron light nonproportionality has some impact on fast electrons as well. Although it has been historically assumed that the scintillation light produced by fast electrons is proportional to the energy deposited in an organic scintillator~\cite{BirksTheoryAndPractice1964}, recent measurements demonstrate electron light yield nonproportionality as high as several MeV for plastic and liquid organic scintillators~\cite{Nassalski2008,Swiderski2012}. Even in a proportional regime, fast electrons lose energy along their track by exciting and ionizing the scintillating medium, and the light yield is the \textit{integrated} light output per electron energy deposited. The increased specific energy loss (as the electron slows) contributes nonproportionally to the total light yield, though the relative contribution decreases with increasing electron energy. Thus, electron recoils serve as a poor reference for unquenched ions in organic scintillators. Quenching factors should therefore be defined relative to the light produced by an unquenched, weakly ionizing particle.  

\begin{figure}
	\center
	\includegraphics[width=0.49\textwidth]{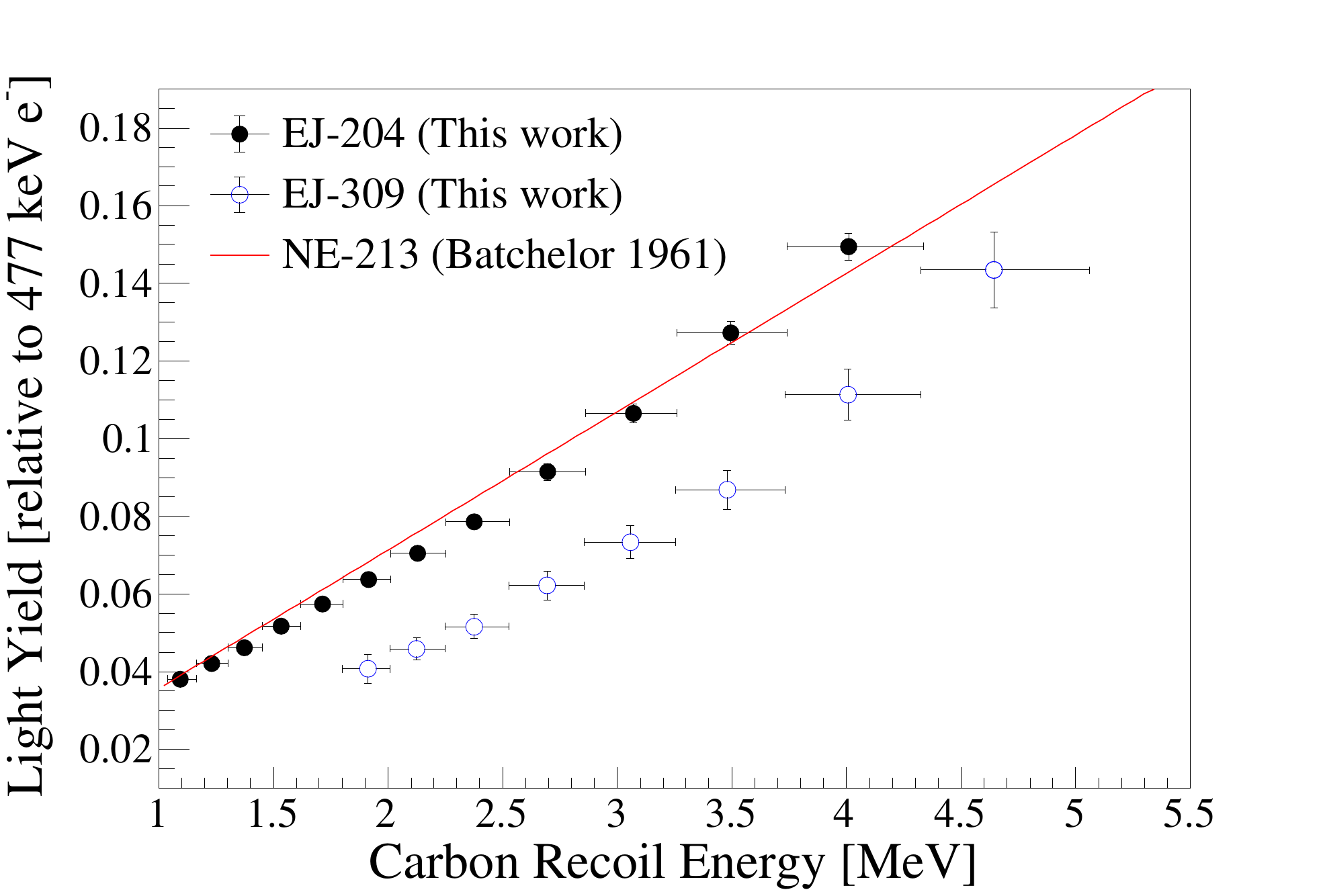}
	\caption{Carbon light yield of EJ-204 (filled circles) and EJ-309 (open blue squares) along with an empirically determined functional form for NE-213 (Batchelor 1961:~\cite{Batchelor1961}). \label{Batchelor}}
\end{figure}

Figure~\ref{Batchelor} shows the carbon light yield relations of EJ-309 and EJ-204 measured in this work, along with a functional form provided by Batchelor et al.~\cite{Batchelor1961}: 
\begin{equation}
L=0.017E_{\text{C}},
\end{equation}
where $L$ is the light output in MeVee and $E_{\text{C}}$ is the carbon recoil energy in MeV. The electron-equivalent light unit was converted to the relative light unit used in this work by dividing by 0.477. In the Batchelor et al.\ study, the response of the PSD-capable liquid scintillator, NE-213, was modeled using a Monte Carlo calculation. A linear carbon light yield was assumed with the slope based on calculations using the Birks formula~\cite{Birks1951-spec} and an empirically-determined NE-102 quenching parameter. The estimated carbon light yield from Batchelor et al.\ is in general agreement with the EJ-204 data measured in this work (which shares a common PVT base with NE-102), though differences in the shape suggest disagreement if the data are extrapolated. 

As shown in Fig.~\ref{Batchelor}, the relative carbon light yield of EJ-309 is lower than that of EJ-204, and the fractional difference in the light yield of the two materials decreases with increasing carbon recoil energy. That is, the EJ-309 carbon light yield relative to the light produced by a 477~keV electron is lower than that of EJ-204 by approximately $55\%$ and $33\%$ at 2~MeV and 4~MeV, respectively. This is in contrast to the behavior observed for the relative proton light yield relations of these materials for proton energies of 2 to 4~MeV, where literature measurements demonstrate a relative proton light yield of EJ-309 greater than that of EJ-204 ~\cite{Brown2018,Laplace2020,Laplace2020_EJ309}. In this energy range, excitation densities are over an order of magnitude greater for carbon recoils compared to protons. The difference in proton and carbon quenching in the two materials may be attributed to a number of competing factors. Exciton diffusion is enhanced in liquid scintillators relative to plastics, which increases the probability of bimolecular interactions such as the triplet-triplet annihilation that gives rise to PSD~\cite{Zaitseva2012}. At the same time, bimolecular quenching increases with increasing excitation density~\cite{Christensen2018}. Future work will involve a study of physics-based models of ionization quenching in organic scintillators taking into account the response of proton and carbon recoil nuclei over a broad energy range.  

\section{Summary}
\label{summary}
The EJ-309 and EJ-204 proton and carbon light yield were measured using a double TOF technique over a recoil energy range of approximately 0.3 to 1~MeV and 2 to 5~MeV, respectively for EJ-309 and 0.2 to 0.5~MeV and 1 to 4~MeV, respectively for EJ-204. The proton light yield relations obtained for both media were compared to literature measurements and the results are in good agreement. The EJ-309 carbon light yield compared favorably to a recent measurement from Norsworthy et al.~\cite{Norsworthy2018}, but was approximately 40\% lower than that predicted by the Batchelor formalism~\cite{Batchelor1961}. This work extends knowledge of the EJ-309 carbon light yield and provides the first measurement of the light output for carbon recoils in EJ-204 via an experimental setup that enables an in situ benchmark against proton light yield standards of reference. 

\begin{acknowledgments}
The authors thank the 88-Inch Cyclotron operations and facilities staff for their help in performing these experiments. This work was performed under the auspices of the U.S. Department of Energy by Lawrence Berkeley National Laboratory under Contract DE-AC02-05CH11231. The project was funded by the U.S. Department of Energy, National Nuclear Security Administration, Office of Defense Nuclear Nonproliferation Research and Development (DNN R\&D). This material is based upon work supported in part by the U.S.\ Department of Energy National Nuclear Security Administration through the Nuclear Science and Security Consortium under Award DE-NA0003180 and Lawrence Livermore National Laboratory under Contract DE-AC52-07NA27344. Sandia National Laboratories is a multimission laboratory managed and operated by National Technology and Engineering Solutions of Sandia LLC, a wholly owned subsidiary of Honeywell International Inc., for the U.S. Department of Energy's National Nuclear Security Administration under Contract DE-NA0003525. 
\end{acknowledgments}

\appendix*
\section{LIGHT YIELD DATA}
\label{LYDataPoints}
The proton and carbon light yield data for EJ-309 and EJ-204 are summarized in Tables~\ref{309PLYResultsTable} and~\ref{204PLYResultsTable}, respectively. The asymmetric proton recoil energy bin widths are reflective of the non-uniform distribution of proton energies within a given bin. Covariance matrices (available upon request) were generated using the EJ-309 and EJ-204 Monte Carlo estimation of the systematic uncertainties. As with previous double TOF proton light yield measurements~\cite{BrownThesis,Brown2018}, the uncertainties on the data points are highly correlated. While model parameters may be obtained by fitting the provided data, the covariance matrix should be taken into account to obtain accurate estimates of parameter uncertainties.

\begin{table*}[!htp]
	\caption{Relative proton and carbon light yield data for EJ-309. Recoil energy bin widths are provided as well as the light output uncertainties.}
	\label{309PLYResultsTable}
	\centering
	\renewcommand{\arraystretch}{1.3}
	\setlength{\tabcolsep}{8pt}
	\begin{tabular}{cccc}
		\hline
		\hline
		Proton Recoil   &  EJ-309 Light Yield & Carbon Recoil & EJ-309 Light Yield  \\
		Energy [MeV]  &  [rel. 477 keV electron]   & Energy [MeV] &  [rel. 477 keV electron] \\
		\hline
		0.340$_{-0.035}^{+0.036}$ & 0.0779 $\pm$ 0.0060 & 1.91$_{-0.11}^{+0.10}$ & 0.0407 $\pm$ 0.0038 \\
		0.419$_{-0.043}^{+0.044}$ & 0.1084 $\pm$ 0.0082 & 2.12$_{-0.12}^{+0.12}$ & 0.0459 $\pm$ 0.0028 \\
		0.515$_{-0.052}^{+0.053}$ & 0.1497 $\pm$ 0.0108 & 2.37$_{-0.13}^{+0.15}$ & 0.0516 $\pm$ 0.0031 \\
		0.630$_{-0.062}^{+0.066}$ & 0.2081 $\pm$ 0.0153 & 2.69$_{-0.17}^{+0.16}$ & 0.0621 $\pm$ 0.0037 \\
		0.772$_{-0.077}^{+0.077}$ & 0.2860 $\pm$ 0.0201 & 3.05$_{-0.20}^{+0.20}$ & 0.0734 $\pm$ 0.0043 \\
		0.927$_{-0.078}^{+0.107}$ & 0.3682 $\pm$ 0.0234 & 3.47$_{-0.23}^{+0.25}$ & 0.0868 $\pm$ 0.0051 \\
		&                     & 4.00$_{-0.27}^{+0.32}$ & 0.1113 $\pm$ 0.0066 \\ 
		&                     & 4.64$_{-0.32}^{+0.42}$ & 0.1434 $\pm$ 0.0097 \\ 
		\hline
		\hline
	\end{tabular}
\end{table*}

\begin{table*}[!ht]
	\caption{Relative proton and carbon light yield data for EJ-204. Recoil energy bin widths are provided as well as the light output uncertainties.}
	\label{204PLYResultsTable}
	\centering
	\renewcommand{\arraystretch}{1.3}
	\setlength{\tabcolsep}{8pt}
	\begin{tabular}{cccc}
		\hline
		\hline
		Proton Recoil   &  EJ-204 Light Yield & Carbon Recoil & EJ-204 Light Yield  \\
		Energy [MeV]  &  [rel. 477 keV electron]   & Energy [MeV] &  [rel. 477 keV electron] \\
		\hline
		0.223$_{-0.023}^{+0.025}$ & 0.0405 $\pm$ 0.0018 & 1.09$_{-0.06}^{+0.07}$ & 0.0381 $\pm$ 0.0009 \\
		0.276$_{-0.028}^{+0.030}$ & 0.0531 $\pm$ 0.0027 & 1.23$_{-0.07}^{+0.07}$ & 0.0421 $\pm$ 0.0010 \\
		0.339$_{-0.034}^{+0.037}$ & 0.0712 $\pm$ 0.0032 & 1.37$_{-0.07}^{+0.08}$ & 0.0462 $\pm$ 0.0011 \\
		0.419$_{-0.043}^{+0.044}$ & 0.0975 $\pm$ 0.0050 & 1.54$_{-0.08}^{+0.08}$ & 0.0517 $\pm$ 0.0012 \\
		0.512$_{-0.049}^{+0.056}$ & 0.1298 $\pm$ 0.0057 & 1.71$_{-0.10}^{+0.09}$ & 0.0574 $\pm$ 0.0014 \\
		&                     & 1.92$_{-0.11}^{+0.10}$ & 0.0638 $\pm$ 0.0015 \\
		&                     & 2.13$_{-0.12}^{+0.12}$ & 0.0705 $\pm$ 0.0016 \\
		&                     & 2.38$_{-0.12}^{+0.16}$ & 0.0786 $\pm$ 0.0019 \\
		&                     & 2.70$_{-0.16}^{+0.17}$ & 0.0915 $\pm$ 0.0022 \\
		&                     & 3.07$_{-0.21}^{+0.19}$ & 0.1066 $\pm$ 0.0024 \\
		&                     & 3.50$_{-0.24}^{+0.24}$ & 0.1272 $\pm$ 0.0030 \\
		&                     & 4.01$_{-0.27}^{+0.32}$ & 0.1495 $\pm$ 0.0034 \\
		\hline
		\hline
	\end{tabular}
\end{table*}

\pagebreak

\bibliographystyle{apsrev4-1} 
\bibliography{./CarbonLY}

\end{document}